\documentclass[12pt,osajnl2,preprint,showpacs,superscriptaddress]{revtex4}  
\usepackage[draft]{hyperref} 
\usepackage{graphics}

\begin{document}

\title{SUDDEN DEATH AND BIRTH OF ENTANGLEMENT EFFECTS FOR KERR-NONLINEAR COUPLER
}

\author{A. Kowalewska-Kud{\l}aszyk}
\affiliation{Nonlinear Optics Division, Department of Physics, Adam Mickiewicz
 University, Umultowska 85, 61-614 Pozna\'n, Poland}
\email{annakow@amu.edu.pl}
\author{W. Leo\'nski}
\affiliation{Nonlinear Optics Division, Department of Physics, Adam Mickiewicz
 University, Umultowska 85, 61-614 Pozna\'n, Poland}

\begin{abstract}
We analyse the entanglement dynamics in a nonlinear Kerr-like coupler interacting with external environment. Whenever the reservoir is in a thermal vacuum state the entanglement (measured by concurrence for a two-qubit system) exhibits regular oscillations of decreasing amplitude. In contrast, for thermal reservoirs we can observe dark periods in concurrence oscillations (which can be called a "sudden death" of the entanglement) and the entanglement rebuild (which can be named the "sudden birth" of entanglement). We show that these features can be observed when we deal with 2-qubit system as well as qubit-qutrit system.
\end{abstract}


\maketitle

\section{Introduction}
One of the fundamental areas of interest in quantum information theory concerns problems of generation of entanglement  in various quantum systems and the influence of external environment on disentanglement processes. As decoherence processes in entanglement evolution are unavoidable consequences of the system-environment interactions, one should consider this problem 
seriously in any real physical realisation of quantum computing devices. 
Especially interesting in this context are recently described phenomena of sudden death of entanglement \cite{ZHHH01,YE04} which consist in entanglement disappearance in finite time rather than exponential decoherence decay of an individual qubit, and a sudden birth of entanglement \cite{FT06} characterised by the reappearance of entanglement after its death. Both these phenomena are caused by the interaction of the pair of entangled qubits with environment. For the systems described by Yu and Eberly \cite{YE04} and also by Ficek and Tana\'s \cite{FT06} an entangled pair was formed by the two 2-level atoms. The initially entangled atoms were located in separate cavities \cite{YE04} (initially in the vacuum state) and had no direct interactions with each other during the evolution. A finite-time decoherence of the system having non-local properties is therefore caused by the spontaneous emission processes.
Similarly, a sudden birth of entanglement in the system considered in \cite{FT06} (composed of two initially entangled 2-level atoms inside a single cavity in a vacuum state) is a consequence of interaction with the environment. The particularly, crucial role plays the formation of collective states of these two atoms followed by collective damping, which cannot be neglected whenever the atoms are close to each other.
Moreover, a phenomenon of delayed birth of entanglement in the previously mentioned system was described in \cite{FT08}.
It was also shown that the thermal reservoir \cite{ILZ07,QJ08} or a reservoir in a squeezed vacuum state \cite{ILZ07} always causes sudden death of entanglement of a two-qubit system (two 2-level atoms) for certain classes of initial entangled states. 

The decoherence process of entanglement between two harmonic oscillators coupled to the reservoir with the spectral density of the Ohmic character was considered in \cite{PR08}. Various kinds of disentanglement evolution, depending on the characteristic features of that environment were considered. The both phases of evolution: (i) sudden death, (ii) sudden death and entanglement revival were identified as well as the usually observed phase of no sudden death behaviour.

Some systems consisting of harmonic oscillators were also previously described in order to identify their various disentanglement properties.
Within the Markovian \cite{SIPS04} and for some cases under non-Markovian \cite{MOP07} approximations there is a possibility for obtaining a sudden death phenomenon, or for some cases in a non-Markovian reservoir the entanglement persists  for longer times than in a Markovian reservoir. Under some conditions, the final state of the initially entangled pair can stay entangled (but with lower value of entanglement).
It is also possible to obtain entanglement of the initially disentangled pair via the interaction with a Markovian reservoir \cite{BF06}.

In the present paper we will focus on the problem of evolution of entanglement generated between two anharmonic oscillators placed inside the two-mode cavity. We will show that it is possible to obtain a finite-time disentanglement (a sudden-death) of initially entangled qubits and for some cases also a revival of entanglement --- a sudden-birth. We will consider not only the zero temperature reservoir but the thermal one  as well, and we will show that the latter type of external environment is more suitable for observing the phenomena of sudden death of entanglement and its revival. 

\section{The model}
We deal with the model of Kerr coupler described in \cite{KL06}. It was shown in that paper that when the initial state of the system was an excited one $|2\rangle_a|0\rangle_b$ the system could be used as a generator of maximally entangled states (MES). From the quantum information theory point of view it was a qubit-qutrit system. The MES were generated most efficiently when the damping process of the generated photon states was negligible. The damping process usually causes decoherence in the system. This fact was shown on the basis of analysis of the time dependence of the fidelity  between the quantum state generated and the Bell-like states.

In the present considerations we will focus on the dynamics of the initially generated MES state (in the above
described system) which is then left in a damped cavity. We will show that despite the damping processes the system is able to spontaneously generate entanglement after its sudden vanishing. In a two-atom system such a phenomenon is called a "sudden birth" of entanglement and it was discussed in \cite{FT06}. A "sudden death" of the entanglement between two atoms was studied for example in \cite{YE04-FT08} .
In our case the situation is qualitatively different as we deal with the entanglement between photon states and damping takes place within these photon states.

The system under consideration is described by the following hamiltonian:
\begin{equation}
\hat{H}=\hat{H}_{NL}+\hat{H}_{int}+\hat{H}_{ext}\,\,\,,
\label{heq1}
\end{equation}
where 
\label{leq2}
\begin{eqnarray}
\hat{H}_{NL}&=&\frac{\chi_a}{2}(\hat{a}^\dagger)^2\hat{a}^2+
\frac{\chi_b}{2}(\hat{b}^\dagger)^2\hat{b}^2\,\,\,,\label{eq2a}\\
\hat{H}_{int}&=&\epsilon (\hat{a}^\dagger)^2\hat{b}^2+
\epsilon^* (\hat{b}^\dagger)^2\hat{a}^2\,\,\,,\label{eq2b}\\
\hat{H}_{ext}&=&\alpha \hat{a}^\dagger+{\alpha}^* \hat{a}\,\,\,.\label{eq2c}
\end{eqnarray}
As usual, $\hat{a}(\hat{a}^{\dagger})$ and $\hat{b}(\hat{b}^{\dagger})$ are photon annihilation (creation) operators in modes $a$ and $b$, respectively, $\chi_a(\chi_b)$ are the nonlinearity parameters, $\alpha$ is the strength of the coupling between the external coherent field and the cavity mode $a$, and $\epsilon$ describe the coupling between the two oscillators.
The non-zero nonlinear coupling between the modes of the coupler is responsible for generation of the entanglement between the two-mode photon states. When the initial system's state is an excited two-mode one, for the values of $\epsilon$ and $\alpha$, small when compared with $\chi$, the form of hamiltonian (\ref{eq2b}) indicates that  the states $|2\rangle_a|0\rangle_b$ and $|0\rangle_a|2\rangle_b$ are involved in the system dynamics. Additionally, as a result of the action of the  external field, the state $|1\rangle_a|2\rangle_b$ must be also considered. As it has been proved in \cite{KL06} the maximum entanglement between the states $|2\rangle_a|0\rangle_b$ and $|0\rangle_a|2\rangle_b$ can be formed and the entanglement between $|2\rangle_a|0\rangle_b$ and $|1\rangle_a|2\rangle_b$ states is also possible.

\section{The influence of damping }
As we want to describe the system in a damping environment we should use the density matrix approach. In the standard (Born and Markov) approximations, the adequate master equation can be expressed in the well known form:
\begin{eqnarray}
\frac{d\hat{\rho}}{dt}&=&-\frac{1}{i}\left(\hat{\rho}\hat{H}-\hat{H}\hat{\rho}\right)+\sum\limits_{k=1}^2\left[\hat{C}_{k}\hat{\rho}\hat{C}^{\dagger}_{k}-\frac{1}{2}\left(\hat{C}^{\dagger}_{k}\hat{C}_{k}\hat{\rho}-\hat{\rho}\hat{C}^{\dagger}_{k}\hat{C}_{k}\right)\right]\\
&+&\sum\limits_{k=1}^2\left[\hat{C}_{k_n}^{\dagger}\hat{\rho}\hat{C}_{k_n}+\hat{C}_{k_n}\hat{\rho}\hat{C}_{k_n}^{\dagger}-
\hat{C}_{k_n}^{\dagger}\hat{C}_{k_n}\hat{\rho}-\hat{\rho}\hat{C}_{k_n}\hat{C}_{k_n}^{\dagger}\right]\nonumber
\,\,\, ,
\label{master}
\end{eqnarray}
where the operators $\hat{C}_k$ describe the damping in the modes $a$ and $b$ and are defined as: 
\begin{eqnarray}
\label{eq12}
\hat{C}_{1}&=&\sqrt{2\gamma_{a}}\:\hat{a}\,\,\, ,\nonumber\\
\hat{C}_{2}&=&\sqrt{2\gamma_{b}}\:\hat{b}\,\,\, ,\nonumber\\
\hat{C}_{1_n}&=&\sqrt{\gamma_{a}n_a}\:\hat{a}\,\,\, ,\\
\hat{C}_{2_n}&=&\sqrt{\gamma_{b}n_b}\:\hat{b}\,\,\, .\nonumber
\end{eqnarray}
We have introduced the interaction with thermal baths of non-zero temperature via $\hat{C}_{1_n}$ and $\hat{C}_{2_n}$ terms, which include the mean photon numbers $n_a$ and $n_b$ of the system's environment, whereas $\hat{C}_{1}$ and $\hat{C}_{2}$ describe interaction with zero temperature bath.

The degree of the entanglement can be described via various parameters. One of them is the concurrence defined for the two qubit system by Wooters \cite{W98}. The value of the concurrence changes from $0$ for completely unentangled states, to $1$ for the maximally entangled ones. This quantity is defined as:
\begin{equation}
C(t)=max\left(\sqrt{\lambda_1}-\sqrt{\lambda_2}-\sqrt{\lambda_3}-\sqrt{\lambda_4},\: 0\right)\,\,\, .
\end{equation}
The parameters $\lambda_i$ are the eigenvalues of the matrix $R$ constructed from the density matrix $\hat{\rho}$ (obtained directly from the master equation) via the relation $R=\hat{\rho}_c\hat{\tilde{\rho}}_c$, where $\hat{\rho}_c$ is the density matrix operator for 2-qubit system, and $\hat{\tilde{\rho}}_c$ can be obtained from the relation:
\begin{equation}
\hat{\tilde{\rho}}_c=\sigma_y\otimes\sigma_y\hat{\rho}^\ast_c\sigma_y\otimes\sigma_y\,\,\, .
\end{equation}
Appearing here $\sigma_y$, is the well known $2\times 2$ Pauli matrix. 

While for the two-qubit system the definition of the concurrence can be applied straightforwardly,
for a qubit-qutrit system there is a need to extract from the system's density matrix a matrix for a two-qubit system using a projection operator according to \cite{LX05}. Thus, we get the reduced density operator $\hat{\rho}_c$ given by:
\begin{equation}
\hat{\rho}_c={\Pi_{0,2}\otimes\Pi_{0,2}\hat{\rho}\Pi_{0,2}\otimes\Pi_{0,2}} 
\,\,\, ,
\label{eq5}
\end{equation}
where
\begin{equation}
\Pi_{0,2}\otimes\Pi_{0,2}\equiv(\left|0\right>\left<0\right|
+\left|2\right>\left<2\right|)\otimes(\left|0\right>\left<0\right|
+\left|2\right>\left<2\right|)\,\, .
\end{equation}
In such a way we will investigate the entanglement between the states for a two-qubit case via the inspection of the formation of the two Bell-like states:
\begin{eqnarray}
\label{eq7}
\left|B_1\right>&=&\frac{1}{\sqrt{2}}\left(\left|2\right>_a\left|0\right>_b
+i\,\left|0\right>_a\left|2\right>_b\right)\,\,\,, \nonumber \\
\left|B_2\right>&=&\frac{1}{\sqrt{2}}\left(\left|2\right>_a\left|0\right>_b
-i\,\left|0\right>_a\left|2\right>_b\right)\,\,\, .
\end{eqnarray}
One should remember that apart from these two states there is also a possibility of formation of other entangled states, which in general influence the total amount of entanglement in the system. 

We will analyse the process of sudden disentanglement and the possible reappearance of entanglement in two different situations. At first, we assume that the coupler initially prepared in an entangled state is located inside a cavity and we can ignore the external coherent field ($\alpha=0$). Then in the second case, we can allow the external field to interact with a coupler -- in such a situation we supply the coupler with energy, which leaks through the cavity mirrors (one should remember that it is not the external field that produces the entanglement between the coupler parts).

\section{System without excitation ($\alpha=0$)}
Before we deal with the problem of entanglement in a qutrit-qubit system, first we
will focus on the situation when the interaction with external coherent field is not present ($\alpha=0$). Physically, it corresponds to the faster (than in the $\alpha\neq 0$ case) leakage of the energy from a coupler system through the cavity mirrors, but moreover, it means that there are only a few states involved in the dynamics -- for the case without damping there are the states $|0\rangle|2\rangle$ and $|2\rangle|0\rangle$, whereas, when the damping is present we also have to deal with these states:$|0\rangle|0\rangle$, $|0\rangle|1\rangle$, $|0\rangle|2\rangle$, $|1\rangle|0\rangle$, and $|2\rangle|0\rangle$. 

When we include the damping processes, for the initially entangled state and for $n_a=n_b=0$ we can observe the continuous process of concurrence $C(t)$ oscillations with decreasing amplitude as expected -- see Fig1. We can identify this decoherence behaviour as a non-sudden death process and it is preserved for the vacuum reservoir case only. The whole population is finally transferred to the product state $\frac{1}{\sqrt{2}}|0\rangle_a\otimes\left(|0\rangle + |1\rangle\right)_b$.

For the thermal reservoirs we can observe that the decoherence changes in a different way (see Fig.2).We cannot observe the continuous and periodical vanish of the concurrence amplitude any more. The inset of Fig.2 presents the part of the concurrence evolution
for various values of $n_a$ and $n_b$. Whenever in one of the bath modes (or in both of them) there is a  different from zero mean number of thermal photons we can observe some gaps during the concurrence evolution. The concurrence may completely vanish --- meaning that the states considered are not entangled any more ---  and after some time spontaneously appear again --- the entanglement between the two states considered is rebuild due to the interaction between the Kerr-coupler and the thermal reservoir.

Moreover, we have analysed the process of disappearance and revivals of entanglement for reservoirs with various values of $n_a$ and $n_b$. To simplify the numerical calculations, we have used greater value of the damping constants $\gamma_a$ and $\gamma_b$. For this case we can see that indeed, it is the interaction between the system and the thermal reservoir that causes a sudden disentanglement in the system, and moreover, it may cause a revival of the entanglement after that death. It can be seen from Fig.3 showing  maps of the concurrence for various numbers of photons. The plots are generated in such a way that dark areas correspond to $C=0$ value and the bright ones to any other (higher than zero) values of concurrence.
The individual thin dark lines appearing in the maps reflect to the fact that $C(t)$ during the oscillations reaches zero from time to time (in such a way as in Fig.1 for example), and as such it is not interesting for our considerations. Only dark areas surviving for some time indicate that a sudden death has occurred. Any bright areas after that, indicate a revival of the entanglement in the system. 
We can inspect this fact in more details analysing Fig.4 showing the concurrence for $n_b=3$ and for various $n_a$; with increasing values of $n_a$ and $n_b$ the time between the death and sudden birth of the entanglement becomes longer.

\section{System excited by an external field ($\alpha\neq 0$)}
The second situation we want to focus on is that, when the external coherent field is switched on. This field's role is to preserve the energy in the system composed of two oscillators and the cavity, and to diminish the role of the leakage of energy through the cavity mirrors. One should remember that it is not the external field that causes the entanglement in the system. The entanglement arises due to the nonlinear interaction between the oscillators, which is present during the whole process \cite{KL06}. 
An external coherent field apart from reducing the role of damping is also responsible for engaging other two-mode states in the dynamics. The entanglement in the coupler system without damping appears as a result of formation of the two Bell-like states (\ref{eq7}) and additionally, an entangled state of the form:
\begin{equation}
\label{eq8}
\left|B_3\right>=\frac{1}{\sqrt{2}}\left(\left|2\right>_a\left|0\right>_b
+\,\left|1\right>_a\left|2\right>_b\right)\,\,\, .
\end{equation}
These states are formed alternately and the details can be found in \cite{KL06}. When the coupler interacts with the reservoir, the whole population is finally transferred to the product state $|P\rangle=\frac{1}{\sqrt{2}}\left(|0\rangle+|1\rangle\right)_a\otimes|0\rangle_b$ (in the time which depends on the damping introduced via the transparence of the cavity walls).
Intermediate concurrence changes may appear, especially for the thermal reservoirs, and they may reveal previously mentioned features of sudden "death" and sudden "birth" of the entanglement, and in fact they can be even enhanced as the coupler is continuously supplied with the energy. 

As the reservoir is assumed to be zero temperature bath (in both modes $a$ and $b$) we can observe the concurrence changes of a slightly different character. The two types of concurrence maxima appear --- see the inset $a_1$ at Fig.5. These are two successive maxima of larger amplitude (they correspond to the formation of the Bell-like states (\ref{eq7})) and between the two subsequent pairs of these maxima there are groups of maxima with significantly smaller amplitude (they are the effect of formation other entangled states --- between 
$|2\rangle|0\rangle$ and $|1\rangle|2\rangle$ states).
In this way we deal with the physical system in which the previous entanglement is transferred out from the two-qubit system from time to time, because one of the states of the entangled pair $|2\rangle|0\rangle$ is coupled to the remaining state $|1\rangle|2\rangle$ of the qubit-qutrit system.
One should keep in mind that the system interacts with the thermal bath that can significantly change the concurrence dynamics. This feature cannot be observed when the values of damping constants increase (see the inset ($a_2$) at Fig.5).

We have investigated the influence of the external field strength on the $C(t)$ evolution for a thermal bath with $n_a\neq 0$ and $n_b\neq 0$.
The exemplary map is presented in Fig.6. We can see that even for a relatively small values of $\alpha$ there are explicit areas of no-entanglement and after some time the entanglement appears suddenly again. Moreover, in Fig.7 we can clearly see the effect of the thermal bath on the entanglement formation. For $n_a=n_b=1$, when including the external coupling $\alpha$, we can obtain a significantly enhanced period of disentanglement in the two-qubit system. The period of time when $C(t)=0$ become larger and larger with increasing values of $\alpha$.
For instance, when $\alpha\cong 0.3$  (solid line in Fig.7) between $t=9$ and $t=14$ the system is disentangled, and after $t=14$ the entanglement appears again. 
It means that during that period of time the entanglement is transferred to the state $|B_3\rangle$. Next the population returns to the previously populated two-qubit system.

\section{Conclusions}
The possibility of changing the character of entanglement evolution caused by the external system's environment was considered. In particular, the entanglement obtained in a Kerr-coupler system with nonlinear interaction between its parts, and the process of decoherence caused by a reservoir were analysed. At first the  maximally entangled state in the system was generated, and after that the system started to interact with its external environment. The two cases, namely, that of the cavity interactions with environment of zero temperature ($n_a=n_b=0$) and the second one, when $n_a$ and (or) $n_b$ are different from zero were discussed. The degree of entanglement was studied via the concurrence for an appropriately chosen two-qubit system.
Moreover, analysis of the entanglement evolution for two physically different situations was performed. 
The first one was when the weak interaction with external coherent field was absent during the whole decoherence process and the second one, when such an excitation was present (for recollection --- an external field does not generate an entanglement in a system). We have found for both cases that for a cavity in a vacuum state the process of entanglement decoherence is a continuous and manifested as oscillations of decreasing amplitude. Qualitative changes in the entanglement evolution for non-zero temperature external bath modes were observed. When there are additional photons inside the cavity, it was found possible to observe a sudden disappearance of  entanglement in a two-qubit system, which can be called a "sudden death" of entanglement, and after some time its reappearance, which can be called a "sudden birth" of entanglement.

When $\alpha\neq 0$ it is possible that, due to the interactions with photons inside a cavity, the entanglement can be transferred for some time to another entangled state $|B_3\rangle$ in which one of the two-qubit states is involved, and after some time, it returns to the system.

For the case when $\alpha=0$ there is no additional coupling to $|B_3\rangle$ state. Different from zero number of photons in thermal bath leads to population of the other two-mode coupler states which (due to the interactions between the nonlinear oscillators) during the evolution change the population of the two-qubit system described.
In such a way the interactions with these additional two-mode states (and between them) lead to the population transfers from them to the Bell $|B_1\rangle$ and $|B_2\rangle$ states, causing the revival of the entanglement in the system after its disappearrance. 

\begin{acknowledgments}
This work was supported by the Polish research network LFPPI.
\end{acknowledgments}


\newpage
\section{Figure Captions}
\hspace{1cm}\vspace{-1.5cm}\\
\noindent
{\bf FIG. 1:} The concurrence $C(t)$ for the coupler initially prepared in a MES state and later left inside the cavity for $n_a=n_b=0$. $\chi_a=\chi_b=\chi=25$, $\gamma_a=\gamma_b=0.001$, $\alpha=0$ and $\epsilon=\pi/\chi$. The inset presents the same but in a different time scale. Time is scaled in $1/\chi$ units.

\noindent
{\bf FIG. 2:} The concurrence $C(t)$ for the coupler initially prepared in a MES state and later left inside the cavity for various values of $n_a$ and $n_b$ and for $\gamma_a=\gamma_b=0.01$. The remaining parameters are chosen the same as in Fig.1. The inset presents the same but in a different time scale.

\noindent
{\bf FIG. 3}
The maps showing the dependence of $C(t)$ on time and value of $n_a$ for various values of $n_b$. The other parameters are the same as in Fig.2. Black areas correspond to $C(t)=0$, whereas the white regions to the positive values of $C(t)$.

\noindent
{\bf FIG. 4:}
The concurrence $C(t)$ for the coupler initially prepared in a MES state and later left inside the cavity for $n_b=3$ and various values of $n_a$. The remaining parameters are chosen the same as in Fig.3.

\noindent
{\bf FIG. 5:} The concurrence $C(t)$ for the coupler initially prepared in a MES state and after that left inside the cavity for $n_a=n_b=0$. The parameters $\chi_a=\chi_b=\chi=25$, $\gamma_a=\gamma_b=0.001$, $\alpha=\pi/\chi$ and $\epsilon=\pi/\chi$. The inset $(a_1)$ shows the same but in a different time scale, whereas ($a_2$) is plotted for 
$\gamma_a=\gamma_b=0.01$.

\noindent
{\bf FIG. 6}
The maps showing concurrence $C(t)$ for various moments of time and values of external pumping $\alpha$. The initial state is $|B_1\rangle$, $n_a=n_b=1$. The remaining parameters are chosen the same as in Fig.2.

\noindent
{\bf FIG. 7:} The concurrence $C(t)$ for the coupler initially prepared in a MES state. The other parameters are: $\chi_a=\chi_b=\chi=25$, $\gamma_a=\gamma_b=0.01$, $\epsilon=\pi/\chi$ and $n_a=n_b=1$.

 \begin{figure}[p]
\resizebox{18cm}{10cm}
                {\includegraphics{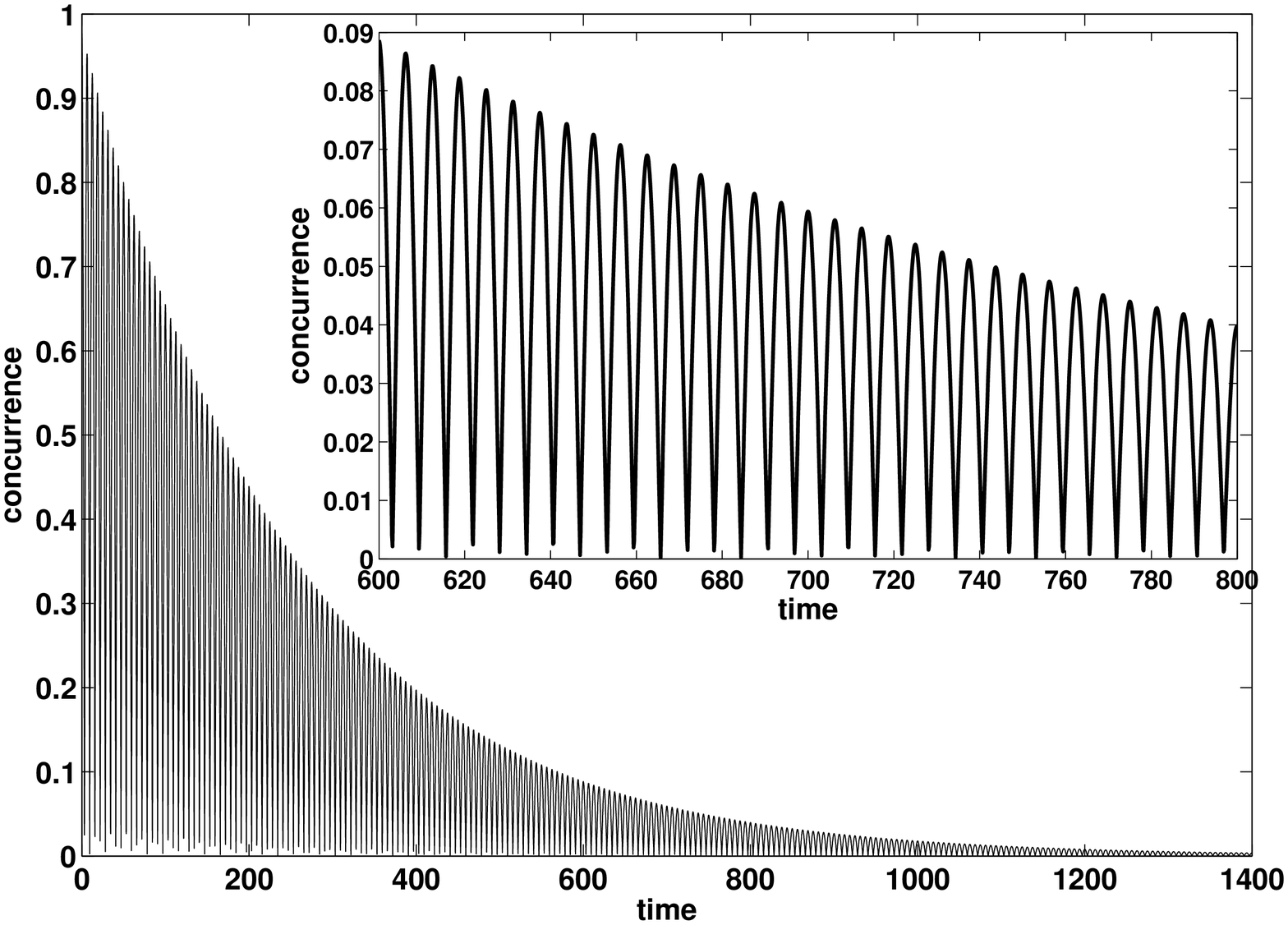}}
\caption{}
\end{figure}

 \begin{figure}
\resizebox{18cm}{10cm}
                {\includegraphics{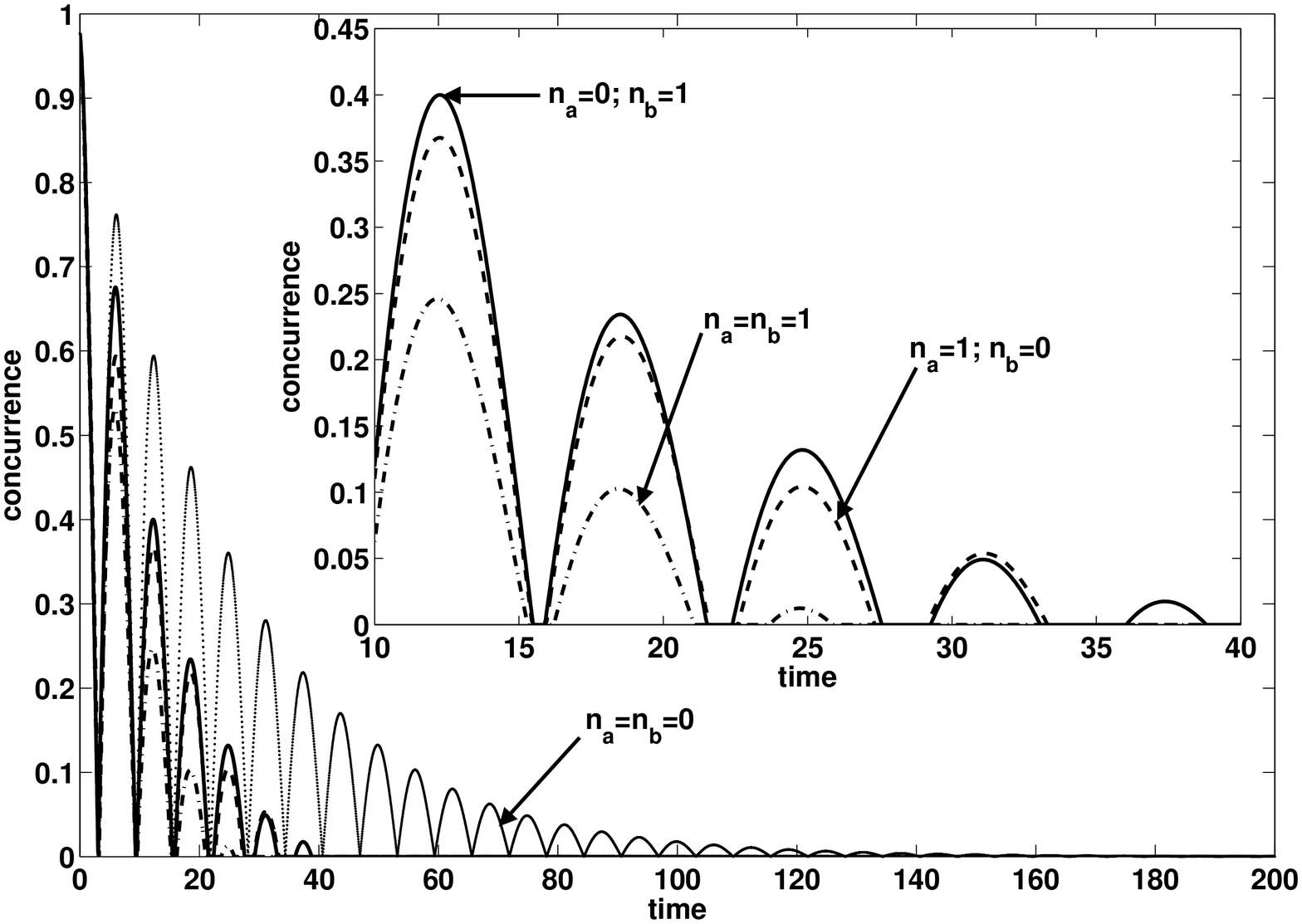}}
\caption{}
\end{figure}

\begin{figure}
\resizebox{8cm}{8cm}
                {\includegraphics{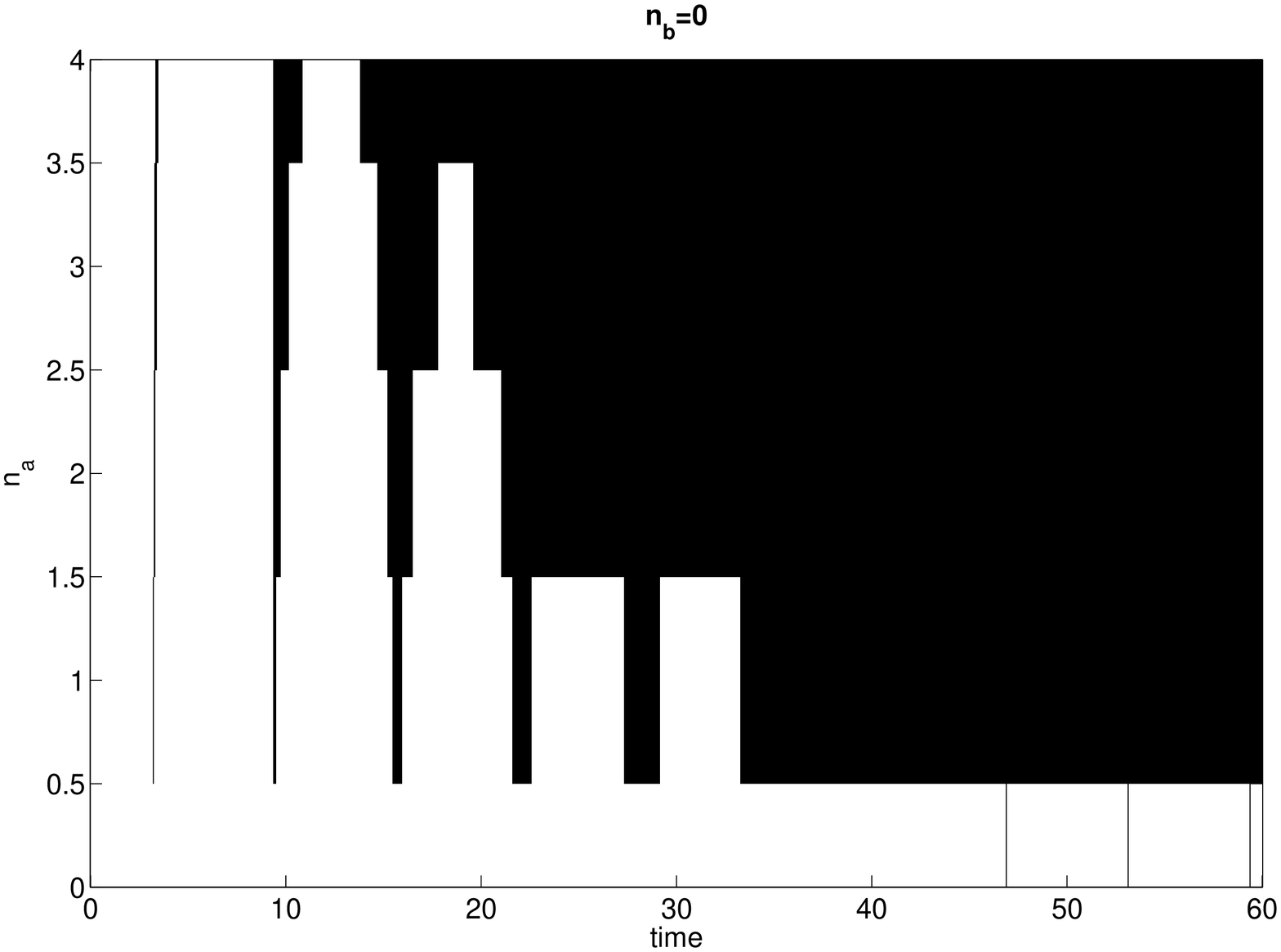}}
\resizebox{8cm}{8cm}
                {\includegraphics{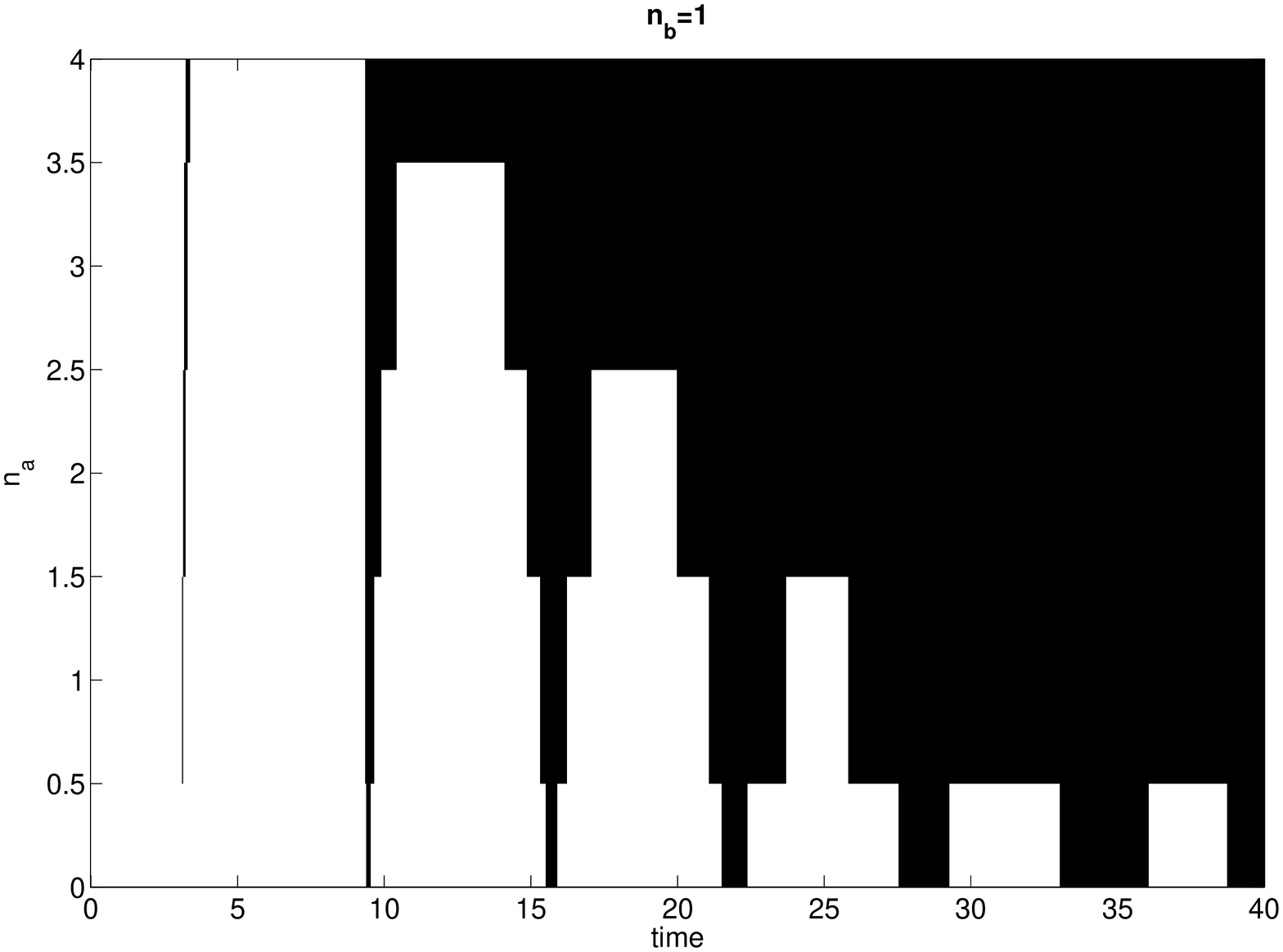}}
\resizebox{8cm}{8cm}
                {\includegraphics{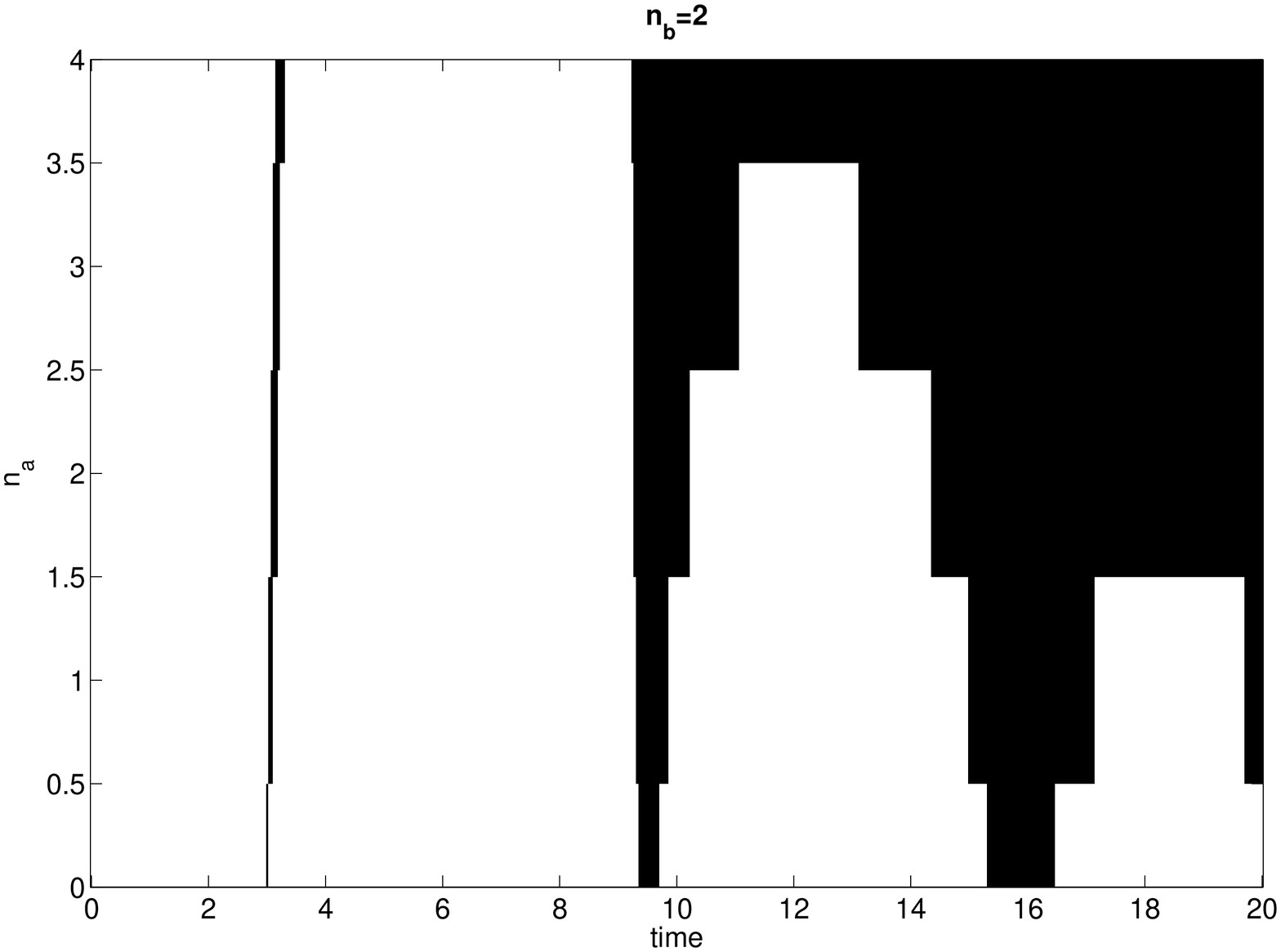}}
\resizebox{8cm}{8cm}
                {\includegraphics{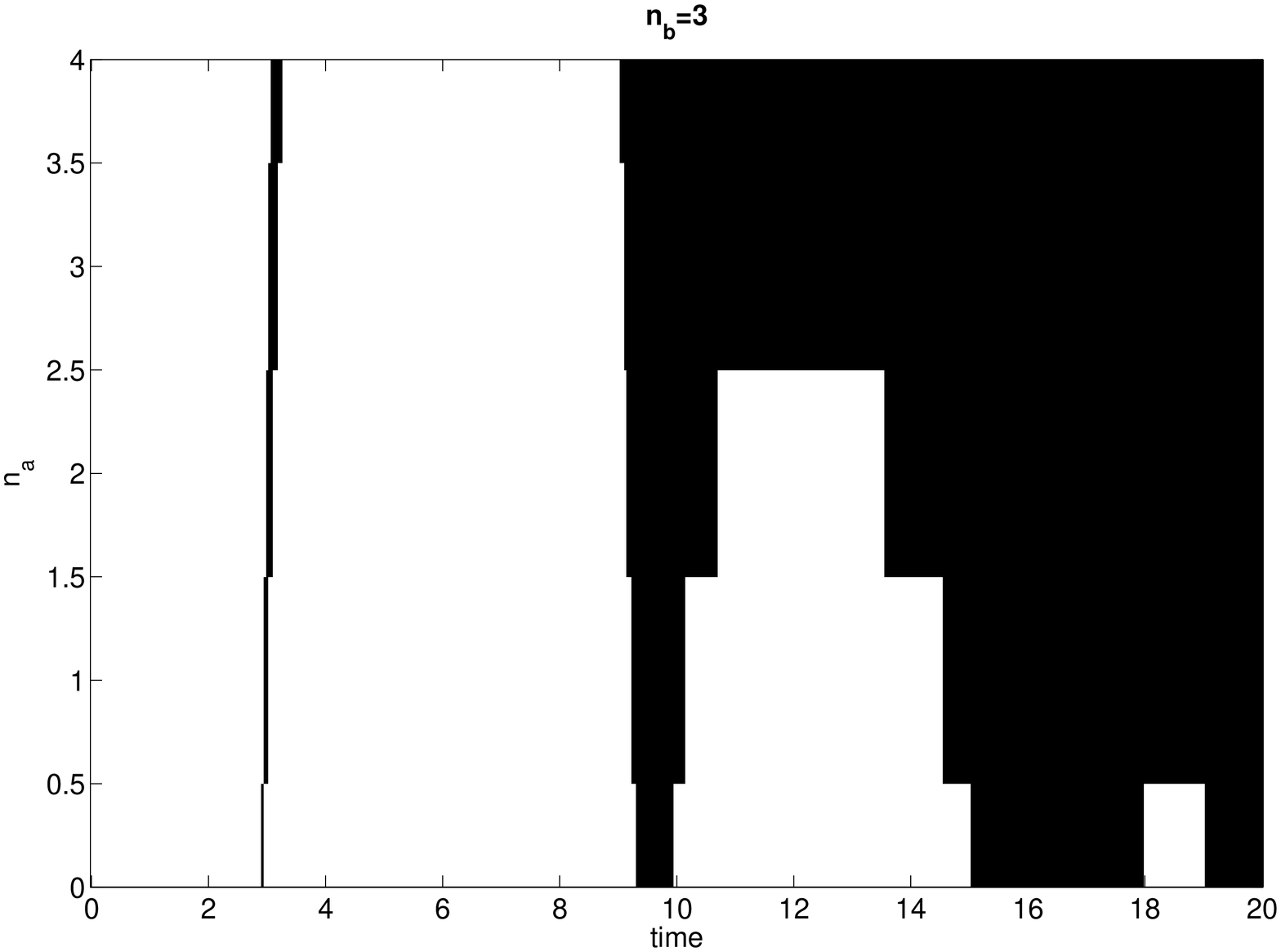}}
\caption{}
\end{figure}

\begin{figure}[p]
\resizebox{18cm}{10cm}
                {\includegraphics{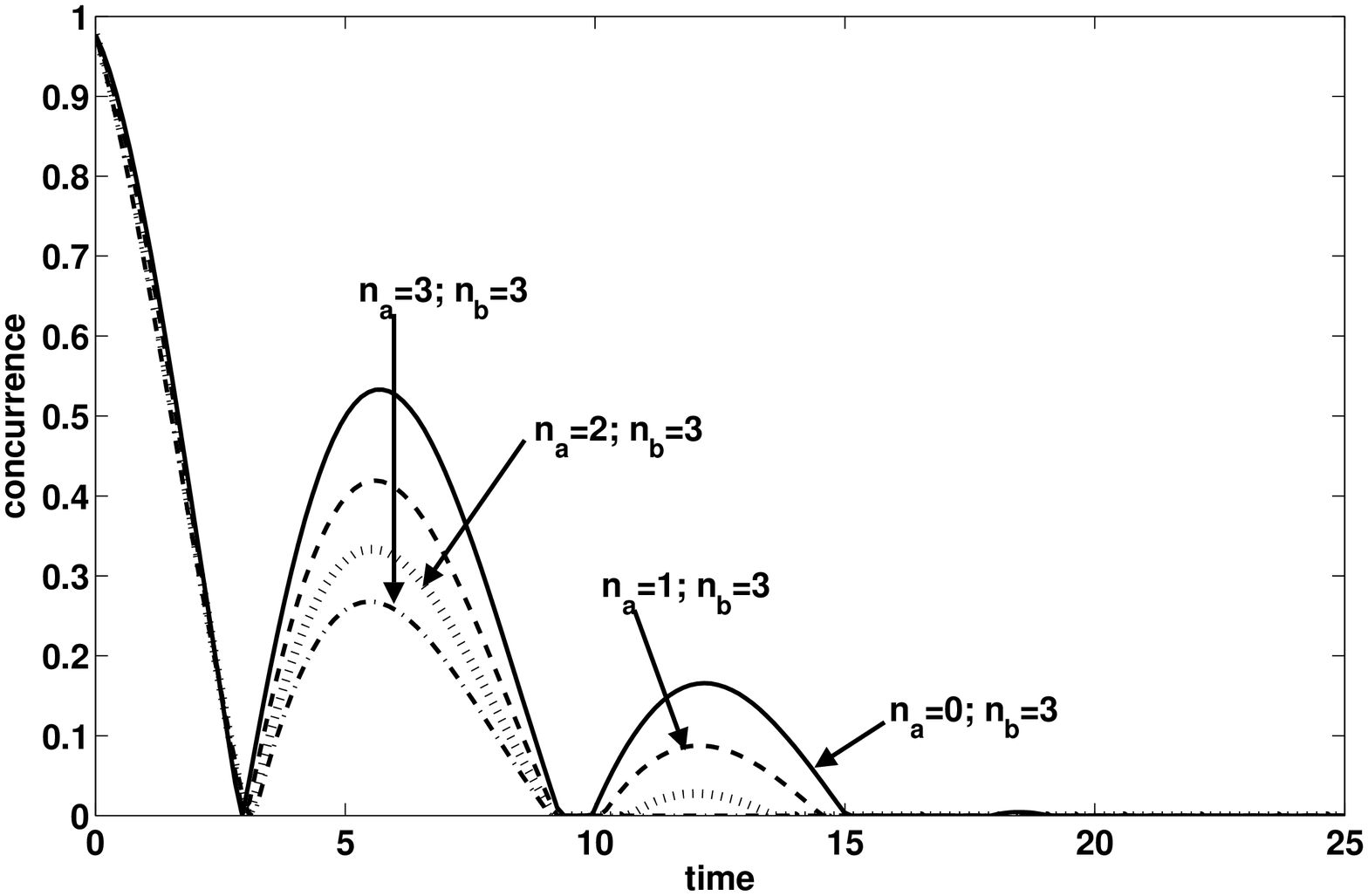}}
\caption{}
\end{figure}

 \begin{figure}
\resizebox{18cm}{10cm}
                {\includegraphics{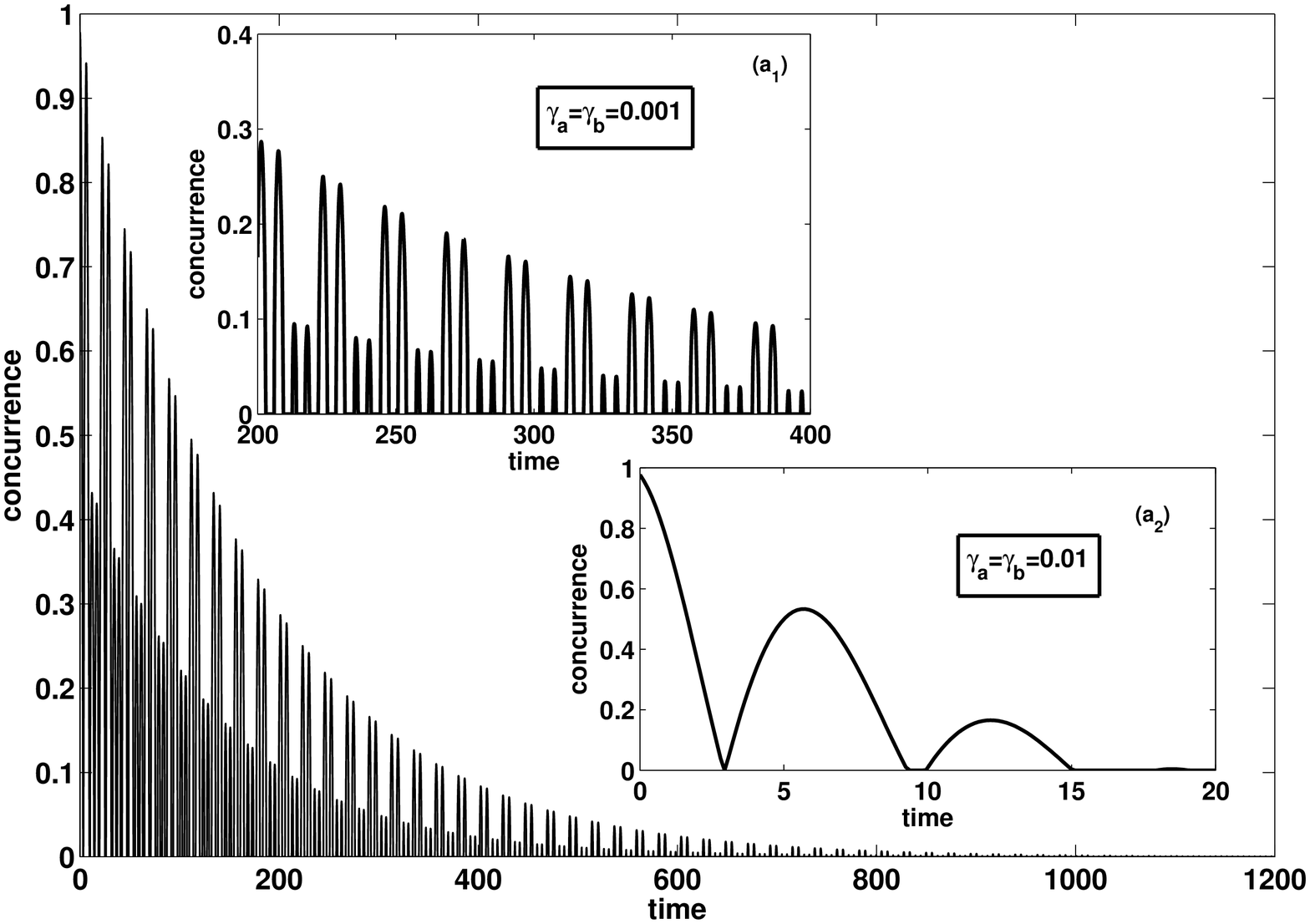}}
\caption{}
\end{figure}

 \begin{figure}
\resizebox{18cm}{10cm}
                {\includegraphics{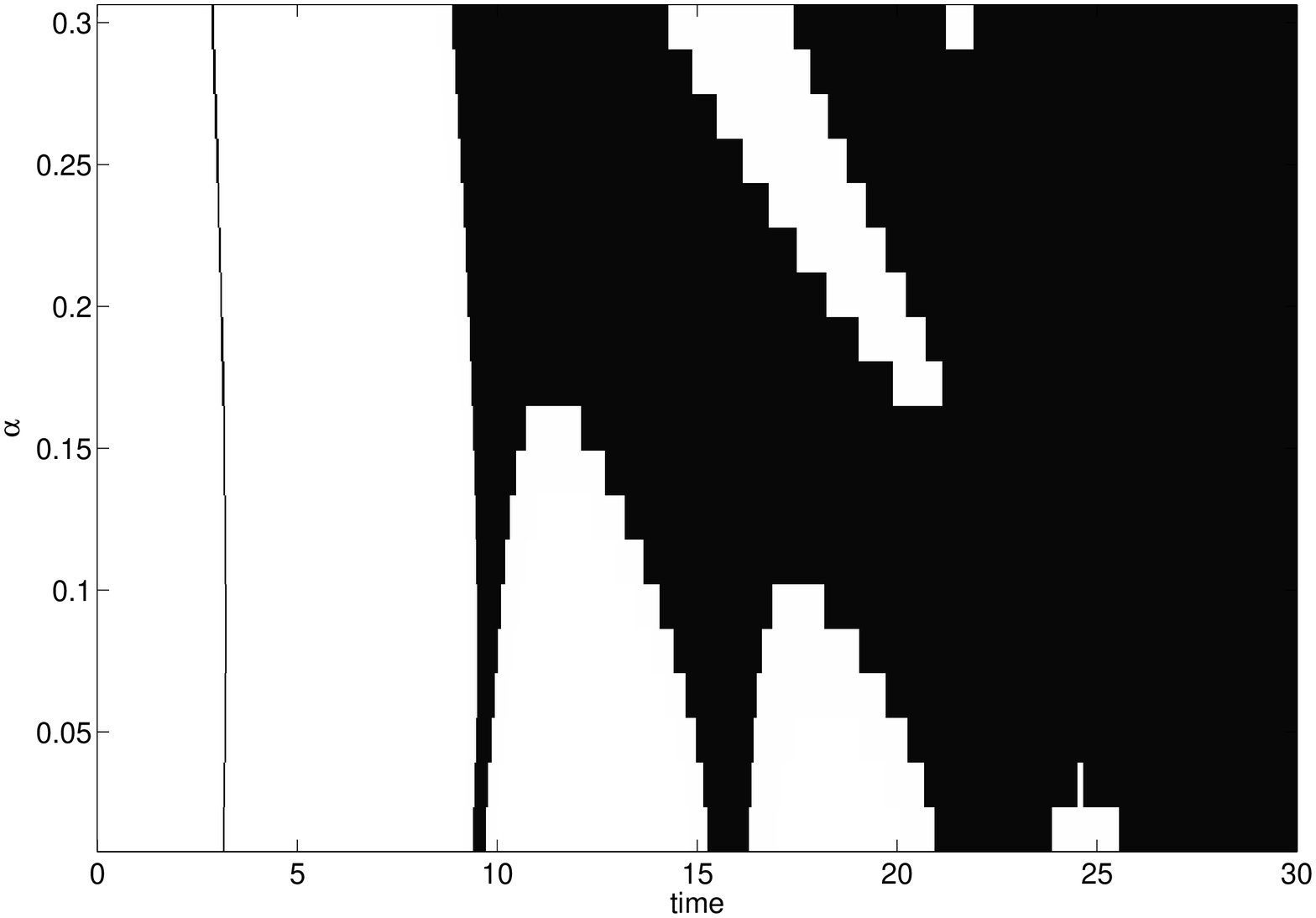}}
\caption{}
\end{figure}

 \begin{figure}
\resizebox{18cm}{10cm}
                {\includegraphics{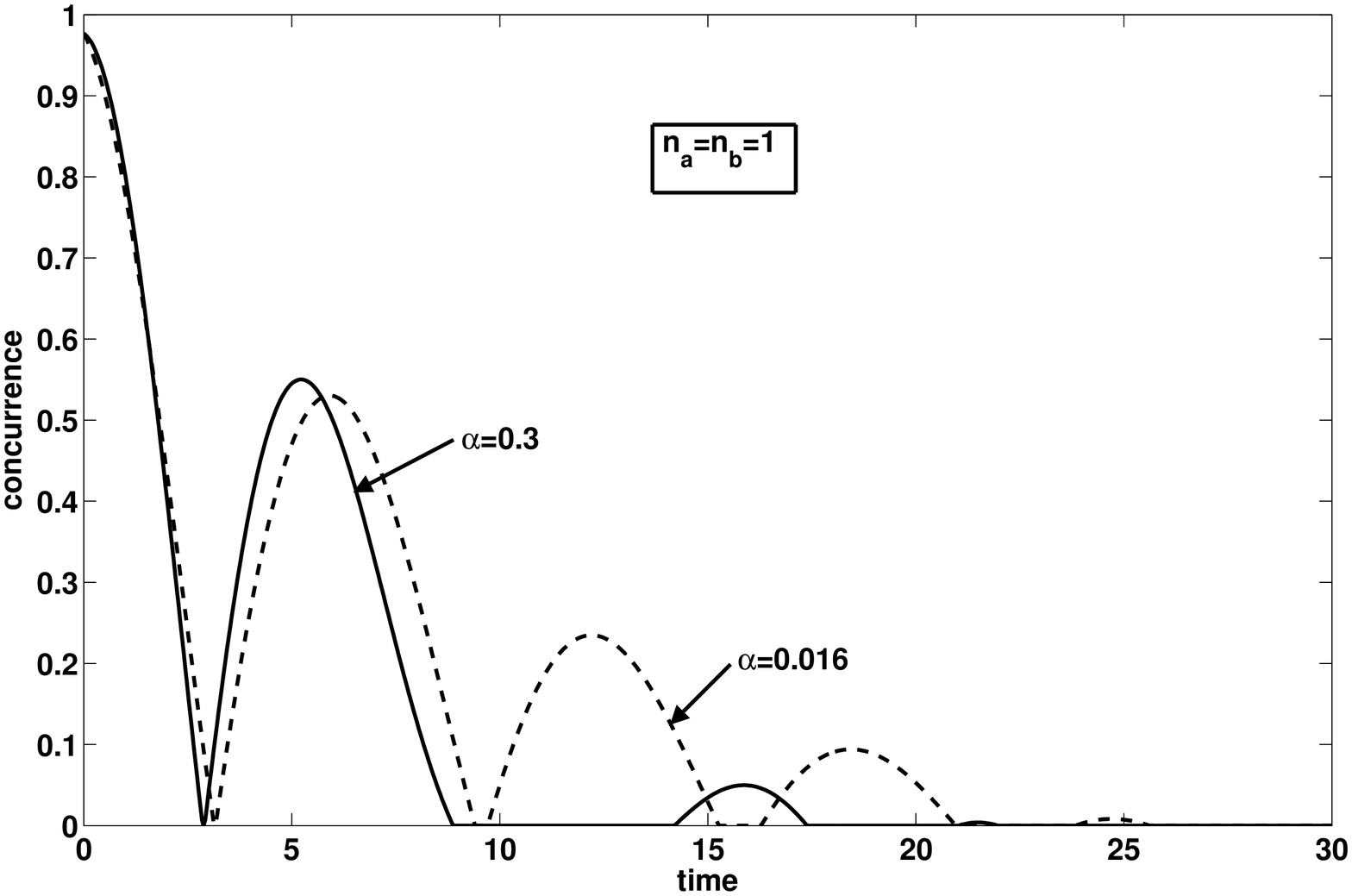}}
\caption{}
\end{figure}

\end{document}